\documentclass[aps,twocolumn,amsmath,amsthm,amssymb,showpacs,preprintnumbers,nofootinbib,prd,superscriptaddress,groupedaddress,10pt]{revtex4-1}

\begin{document}

\title{Mass Ladder Operators from Spacetime Conformal Symmetry}

\author{
Vitor Cardoso$^{1,2}$,
Tsuyoshi Houri$^{3,4}$,
Masashi Kimura$^{1}$
}
\affiliation{${^1}$ CENTRA, Departamento de F\'{\i}sica, Instituto Superior T\'ecnico, Universidade de Lisboa, Avenida~Rovisco Pais 1, 1049 Lisboa, Portugal}
\affiliation{${^2}$ Perimeter Institute for Theoretical Physics, 31 Caroline Street North
Waterloo, Ontario N2L 2Y5, Canada}
\affiliation{${^3}$ Department of Physics, Kobe University, 1-1 Rokkodai, Nada, Kobe, Hyogo 657-8501, Japan}
\affiliation{${^4}$ Department of Physics, Kyoto University, Kitashirakawa, Kyoto 606-8502, Japan}

\date{\today}


\begin{abstract}
Ladder operators can be useful constructs, allowing for unique insight and intuition. In fact, they have played a special role in the development of 
quantum mechanics and field theory. Here, we introduce a novel type of ladder operators, which map a scalar field onto another massive scalar field. 
We construct such operators, in arbitrary dimensions, from closed conformal Killing vector fields, eigenvectors of the Ricci tensor. 
As an example, we explicitly construct these objects in anti-de Sitter spacetime (AdS) and show that they
exist for masses above the Breitenlohner-Freedman (BF) bound. Starting from a regular seed solution of the massive Klein-Gordon equation (KGE), mass ladder operators in AdS allow one to build a variety of regular solutions with varying boundary condition at spatial infinity. We also discuss mass ladder operator in the context of spherical harmonics,
and the relation between supersymmetric quantum mechanics and so-called Aretakis constants in an extremal black hole.
\end{abstract}

\preprint{KOBE-COSMO-17-07}
\preprint{KUNS-2683}

\maketitle

\section{Introduction}
Exactly solvable systems play a crucial role in understanding the physical content of a given theory and
on isolating the main features of certain solutions. Some of such ``golden'' systems, like the hydrogen atom
in quantum mechanics are well-known cornerstones in mathematics, physics and chemistry.
In some cases, it is even possible to relate families of solutions of a given problem without a detailed knowledge of any of its solutions.
One such remarkable technique, in the context of the Schr\"odinger equation, consists in using ladder operators. These enable one to construct algebraically all, or part of, the energy eigenvalues and eigenfunctions. However, finding ladder operators is in general a challenging task, because their relations to the symmetries of the system remain unclear -- usually, therefore, they are called a dynamical symmetry.

Such enterprise is specially relevant within general relativity or the gauge-gravity approach to field theories. 
In such a framework, test fields on curved spacetimes have been repeatedly studied and found -- at least at linearized level -- to inherit the background symmetries; they are thus helpful
in providing a geometric picture of spacetime. Conversely, some techniques were developed which make explicit use of spacetime symmetries to handle these fields.
For example, test fields can be conveniently separated by means of harmonic functions, on spacetimes which are maximally symmetric or contain a maximally symmetric subspace. 
We will focus our attention on the prototypical example of the KGE,
\begin{align}
(\Box - m^2)\Phi = 0\,,
\label{eq:massivescalar}
\end{align}
where $\Box\equiv g^{\mu\nu}\nabla_\mu\nabla_\nu$ is the d'Alembertian and $m$ is a mass parameter.

In this paper, we report that if a background spacetime has a particular conformal symmetry (whose precise conditions will be stated below), there exists a differential operator $D$, called a {\it mass ladder operator}, which maps a solution to the KGE with mass squared $m^2$ into another solution with different mass squared $m^2+\delta m^2$, i.e.,
\begin{align}
(\Box - (m^2+\delta m^2))D\Phi = 0\,,
\label{eq:massivescalar_2}
\end{align}
where $\delta m^2$ is the variation of the mass squared.
Hence, we provide a geometric interpretation to ladder operators in terms of the symmetry of a background spacetime.

Our formulation can be useful in Riemannian geometry, where the KGE
is replaced by the Helmholtz-like equation, {\it i.e.}, the eigenvalue equation for the Laplacian,
\begin{equation}
(\Delta-\lambda)\Phi = 0\,,
\end{equation}
where $\Delta\equiv g^{\mu\nu}\nabla_\mu\nabla_\nu$ is the Laplacian and $\lambda$ is the eigenvalue of the Laplacian. For example, when we 
consider the Laplacian on $S^2$, we obtain ladder operators which change the azimuthal quantum number~\cite{Higuchi:1986wu,Gelderen:1998}. The reason why such ladder operators exist on $S^2$ has never been explained in terms of conformal symmetry. In our formulation, we can explicitly construct the ladder operators from conformal Killing vectors on $S^2$.
\section{Mass ladder operators for scalars}
Conformal symmetry of an $n$-dimensional spacetime $(M,g_{\mu\nu})$ is defined by the invariance for a metric $g_{\mu\nu}$ under the conformal transformation $g_{\mu\nu}\to g'_{\mu\nu}=\exp(2Q)g_{\mu\nu}$, where $Q$ is a function on $M$. The infinitesimal transformation is described by the conformal Killing equation
\begin{equation}
\nabla_\mu\zeta_\nu+\nabla_\nu\zeta_\mu = 2Qg_{\mu\nu} , \quad
Q = \frac{1}{n} \nabla_\mu\zeta^\mu \,,
\label{CKVequation1}
\end{equation}
where $\zeta^\mu$ is known as a conformal Killing vector, and $Q$ is called the associated function.
In particular, a conformal Killing vector is said to be closed if it satisfies the condition $\nabla_{[\mu}\zeta_{\nu]} = 0$. Hence, $\zeta^\mu$ is a closed conformal Killing vector (CCKV) if it satisfies the equation
\begin{equation}
\nabla_\mu\zeta_\nu = Qg_{\mu\nu} , \quad
Q = \frac{1}{n} \nabla_\mu\zeta^\mu \,. \label{CCKVequation}
\end{equation}
Using this equation, we obtain the following result:(the detailed derivation is shown in the next section).

{\it Suppose that an $n$-dimensional spacetime $(M,g_{\mu\nu})$ admits a CCKV $\zeta^\mu$ satisfying (\ref{CCKVequation}). If $\zeta^\mu$ is an eigenvector of the Ricci tensor with a constant eigenvalue, $R^{\mu}{}_{\nu}\zeta^\nu = \chi(n-1) \zeta^\mu$, then there exists a one-parameter family of mass ladder operators
\begin{equation}
D_k \equiv {\cal L}_\zeta - k Q\,, \label{LADOP}
\end{equation}
where ${\cal L}_\zeta$ denotes the Lie derivative with respect to $\zeta^\mu$,
such that the commutation relation with the d'Alembertian $\square \equiv \nabla^\mu \nabla_\mu$ is given by
\begin{equation}
[\Box, D_k] = \chi(2k+n-2) D_k   + 2Q(\Box +\chi k(k+n-1) )\,,
\label{massladdercommutator1}
\end{equation}
where $k$ is a parameter, and the commutator is considered as acting on a scalar field.}

Since the action of the commutator on a scalar field $\Phi$  leads to
\begin{align}
(\square -  (m^2 + \delta m^2)) D_k \Phi  = D_{k-2}(\square - m^2) \Phi \,,
\label{massladderidentity}
\end{align}
together with $m^2 = - \chi k(k+n-1)$ and $\delta m^2=\chi(2k+n-2)$,
$D_k$ maps a solution $\Phi$ to the KGE with mass squared $m^2$ into another solution $D_k\Phi$ with mass squared $m^2 + \delta m^2=-\chi (k-1)(k+n-2)$. \footnote{If $\Phi$ is a solution to the KGE with a source term $S$, i.e., $(\square - m^2) \Phi = S$, one has $(\square -  (m^2 + \delta m^2)) D_k \Phi  = D_{k-2}S$.} Thus, $D_k$ are mass ladder operators which connect massive solutions to the KGE from the mass squared $m^2$ to $m^2+\delta m^2$. \footnote{If $\chi = 0$, $D_k$ maps massless solutions to massless ones. In that case, we have other ladder operators. See Appendix.\ref{anotherladder} for other constructions of ladder operators.} 
This corresponds in terms of $k$ to shifting $k$ to $k-1$.
It is also found that $D_{-k-n+2}$ maps a solution to the KGE into a solution from the mass squared $m^2 = - \chi (k-1)(k+n-2)$ to $m^2 + \delta m^2 = - \chi k(k+n-1)$.
This corresponds to shifting $k-1$ to $k$.\footnote{
Taking the adjoint of Eq.\eqref{massladderidentity},
(see discussion in Ref.~\cite{Wald:1978vm}),
we obtain $D_k^\dag(\square -  (m^2 + \delta m^2))^\dag = (\square - m^2)^\dag(D_k + 2 Q)^\dag$ 
where $\dag$ means the adjoint operator. Since $\square - m^2$ and $\square -  (m^2 + \delta m^2)$
are self adjoint operators, we can see that $(D_k + 2 Q)^\dag = -D_{-k-n+2}$ shifts 
mass from $m^2 + \delta m^2$ to $m^2$.
}
Since $D_k$ is surjective (or onto) every $k$, all the solutions with mass squared $m^2 + \delta m^2$ can be constructed from the solutions with mass squared $m^2$ (see Appendix.\ref{ontomap}). It should be noted that the operators have the index $k$ and its value must be chosen appropriately depending on the mass of a solution to act.

When $D_k$ connects solutions to the KGE with two real mass squared $m^2$ and $m^2+\delta m^2$, $k$ is required to be real, and the following inequalities
must be satisfied
\begin{eqnarray}
\frac{\chi}{4}(n-1)^2 \leq m^2 ,~\chi<0 \quad  \textrm{or} \quad 
m^2 \leq \frac{\chi}{4}(n-1)^2 ,~\chi>0\,,\nonumber\\
\label{murange}
\end{eqnarray}
where the equality is attained for $k=-(n-1)/2$.
Thus, mass ladder operators can exist only when the value of the mass squared $m^2$ satisfies the above inequalities.
We also notice that for a fixed $m^2$, there are two mass ladder operators $D_{k_\pm}$, where
\begin{equation}
k_\pm = \frac{1-n\pm\sqrt{(n-1)^2-4m^2/\chi}}{2} \,.
\label{eqdefkpm}
\end{equation}
These two operators become a mass raising or mass lowering operator, depending on the value of the mass squared $m^2$. For a negative $\chi$, if $m^2>\chi n(n-2)/4$ they correspond to mass raising and lowering operators, respectively, and otherwise both become mass raising operators. For positive $\chi$, the roles are reversed.

Generally, multiples of the ladder operators can be considered. Since we have [cf.\ (\ref{massladderidentity})]
\begin{eqnarray}
&&(\square -  (k-s)(k+s-1)) D_{k-s}\cdots D_{k-1} D_{k} \Phi  \nonumber\\
&& =D_{k-s-2}\cdots D_{k-3} D_{k-2}(\square -k(k+1)) \Phi\,, \nonumber
\end{eqnarray}
the multiple operator $D_{k-s}\cdots D_{k-1} D_{k}$ can shift the mass squared labeled by $k$ to the one by $k-s$.

Physically important examples are maximally symmetric spacetimes. Actually, we can construct mass ladder operators for the KGEs in such spacetimes.\footnote{
General spacetimes admitting mass ladder operators
are discussed in Appendix.\ref{explicitmetriccckv}.}
If we consider $AdS_n$ spacetime with a cosmological constant $\Lambda = \chi(n-1) < 0$, the first inequality in Eq.\eqref{murange} 
coincides with the condition for the mass above BF bound~\cite{Breitenlohner:1982jf, Breitenlohner:1982bm}.
This means we can define mass ladder operator for the massive scalar with the mass above BF bound.
In $dS_n$ spacetime with a cosmological constant $\Lambda = \chi(n-1) > 0$,
the second inequality in Eq.\eqref{murange} 
is the condition for that the solution of KGE does not have oscillation solution 
for long wave limit (see, e.g., \cite{Akhmedov:2013vka}).

\section{Derivation of mass ladder operators}
\label{seciii}

In $n\geq 2$ dimensions, one can show the following commutation relation 
when acting on a scalar
\begin{equation}
[\square, {\cal L}_\zeta] 
= 2 Q\square - (n-2) (\nabla^\mu Q) \nabla_\mu \,, \label{commutation_rel_CKV}
\end{equation}
where $\zeta^\mu$ is a conformal Killing vector satisfying Eq.\ \eqref{CKVequation1}. If $\zeta^\mu$ is a CCKV, it satisfies Eq.\ (\ref{CCKVequation}). Differentiating this equation, we obtain
\begin{equation}
\nabla_\mu Q = \frac{1}{1-n}R_\mu{}^\nu\zeta_\nu \,. \label{appA_prop_CCKV}
\end{equation}
Assuming in addition that $\zeta^\mu$ is an eigenvector of the Ricci tensor,
\begin{equation}
R^\mu{}_\nu\zeta^\nu = \chi(n-1)\zeta^\mu \,, \label{Ricci_CCKV}
\end{equation}
where $\chi$ is constant\footnote{If a spacetime admits two CCKVs $\zeta_1^\mu$ and $\zeta_2^\mu$ which are respectively eigenvectors of the Ricci tensor
with eigenvalues $\chi_1$ and $\chi_2$, a linear combination is also CCKV, but it is not an eigenvector of the Ricci tensor unless $\chi_1 = \chi_2$.
}, we arrive at the condition that the gradient of the function $Q$ is proportional to $\zeta^\mu$,
\begin{align}
 \nabla_\mu Q = -\chi \zeta_\mu \,. \label{eq_nablaQ}
\end{align}
Under this condition, Eq.(\ref{commutation_rel_CKV}) is
\begin{equation}
[\square, {\cal L}_\zeta]
= 2 Q\square +\chi (n-2) {\cal L}_\zeta \,. \label{com_squareLzeta}
\end{equation}
Furthermore, since Eq.\ \eqref{eq_nablaQ} leads to
\begin{align}
 \square Q  + \chi n Q = 0 \,,
\end{align}
we obtain
\begin{align}
[\square, Q] = -2\chi {\cal L}_\zeta - n\chi Q \,. \label{com_squareQ}
\end{align}
Given Eqs.\ \eqref{com_squareLzeta} and \eqref{com_squareQ}, it is easy to calculate the commutation relation between the d'Alembertian and $D_k$ given by Eq.\ \eqref{LADOP} and then obtain Eq.\ (\ref{massladdercommutator1}).

Suppose there are more than one ladder operators $D_{a,k}={\cal L}_{\zeta_a}-k Q_a$ ($a = 1, 2, \cdots, N$) for the KGE with mass squared $m^2 = -\chi k(k+n-1)$. Then it would be natural to compute the commutation relations between them because one expects that they form a Lie algebra. First, it is important to see that the commutator of CCKVs $\zeta_a^\mu$,
\begin{align}
\xi_{ab}^\mu \equiv [\zeta_a,\zeta_b]^\mu 
=\zeta_{a}^{\nu}\nabla_\nu \zeta_{b}^{\mu} -  \zeta_{b}^{\nu}\nabla_\nu \zeta_{a}^{\mu} \,,
\end{align}
becomes a Killing vector, which satisfies the Killing equation $\nabla^\mu \xi_{ab}^\nu + \nabla^\nu \xi_{ab}^\mu = 0$. Here, we have used the condition that $\zeta_a$ are eigenvectors of the Ricci tensor. Hence we have
\begin{align}
[\hat{H}_k, D_{a,k}] &= \chi (2k+n -2)D_{a,k} + 2 Q_a \hat{H}_k
\\
[\hat{H}_k, {\cal L}_{\xi_{a b}}] &= 0,
\end{align}
where $\hat{H}_k \equiv \square + \chi k(k+n-1)$. The first relation shows that, since $Q_a$ is a function, $D_{a,k}$ becomes a ladder operator only for particular scalar fields $\Phi_k$ obeying the equation $\hat{H}_k \Phi_k = 0$. Since we have $\hat{H}_{k+1}(D_{a,k}\Phi_k) = 0$, one can construct $N$ solutions to the equation $\hat{H}_{k+1}\Phi_{k+1}=0$ from a single $\Phi_k$.
The second relation shows that the Lie derivative along the Killing vector $\xi_{ab}^\mu$ acts on any solution to the equation $\hat{H}_k\Phi =0$ as symmetry.
Since it is also shown that
\begin{align}
-\chi d(\zeta_{a \mu}\zeta_b^\mu) =  d(Q_a Q_b) \,,
\end{align}
we find
\begin{align}
-\chi \zeta_{a \mu}\zeta_b^\mu = Q_a Q_b + C_{ab}
\end{align}
with a constant $C_{ab}$.
Thus the commutation relations between the ladder operators $D_{a,k}$ and the Killing vectors $\xi_{ab}^\mu$ constructed from CCKVs are calculated as
\begin{align}
[D_{a,k}, D_{b,k}] &=  {\cal L}_{\xi_{a b}} \,,
\\
[D_{a,k}, {\cal L}_{\xi_{bc}}] &= 
 C_{ab} D_{c,k} - C_{ac}D_{b,k} \,,
\\
[{\cal L}_{\xi_{a b}}, {\cal L}_{\xi_{c d}}] &= 
C_{ad}  {\cal L}_{\xi_{cb}} 
- C_{bd}  {\cal L}_{ \xi_{ca}}  \notag \\
&~~~~ -  C_{ac } {\cal L}_{ \xi_{db}}  
+ C_{bc} {\cal L}_{ \xi_{da}} \,,
\end{align}
which form a Lie algebra. This implies that solutions to the equation $\hat{H}_k\Phi_k=0$ become the representation of this Lie group. As seen later, this is conformal group in a maximally symmetric spacetime.

\section{Mass ladder operators in AdS}
The metric of the $n$-dimensional anti-de Sitter spacetime ($AdS_n$) in Poincar\'e coordinate is
\begin{align}
ds^2 = \frac{dr^2}{r^2} + r^2 \sum_{A,B = 0}^{n-2} \eta_{AB}dx^A dx^B\,,
\end{align}
where $\eta_{AB} = {\rm diag}[-1,1,\cdots,1]$ is the metric on the $n-1$ dimensional Minkowski spacetime $M^{n-1}$,
and $A,B$ run over $0,1,\cdots,n-2$. 
The massive KGE on this spacetime is
\begin{align}
(r^2 \partial_r^2 + n r \partial_r + r^{-2} \square_{M^{n-1}} - m^2) \Phi = 0\,,
\label{adsnkgeq}
\end{align}
where $\square_{M^{n-1}}$ is d'Alembertian on $M^{n-1}$.
In the $AdS_n$ spacetime, there are $n+1$ CCKVs $\zeta_a^\mu$ which satisfy $\nabla_{\mu} \zeta_{a,\nu} = Q_a g_{\mu \nu}$, where $a$ runs 
over $-1, 0, 1, \cdots, n-1$. Since the Ricci curvature satisfies $R_{\mu \nu} = -(n-1) g_{\mu \nu}$, we have $\chi=-1$, and all the CCKVs are eigenvectors of the Ricci tensor. Thus, from (\ref{LADOP}), we obtain $n+1$ mass ladder operators\footnote{
The map between $k = 1$ and $k = 0$ in $AdS_2$ case 
was partially discussed in~\cite{Cruz:1998vf}.
}
\begin{eqnarray}
D_{-1,k} &=& r^2 \partial_r-k r \,,\\
D_{A,k} &=& x^A r^2 \partial_r + r^{-1} \sum_{B = 0}^{n-2} \eta^{AB}\partial_B-k x^A r \,, \\
D_{n-1,k} &=& (-1+r^2 \sum_{A,B = 0}^{n-2} \eta_{AB}x^A x^B) \partial_r + 2 r^{-1} 
\sum_{A= 0}^{n-2} x^A \partial_A \nonumber\\
& & -k\left(r^{-1} + r \sum_{A,B = 0}^{n-2} \eta_{AB} x^A x^B\right)\,.
\end{eqnarray}
Using global coordinates, we can confirm that these mass ladder operators 
are regular beyond the Poincar\'e horizon. Hence, a regular solution to the KGE can be mapped into another regular solution with different mass (see Appendix.\ref{ladderregularityadsn}).

For simplicity, we discuss how the mass ladder operators act on the solution to the KGE under separation of variables, $\Phi = \alpha(x^A) \tilde{\Phi}(r)$.
Then the KGE reduces to
\begin{align}
&(\Box_{M^{n-1}} - L^2)\alpha = 0, 
\\
& \left[
r^2\frac{\partial^2}{\partial r^2}
+
n r \frac{\partial}{\partial r}
- m^2 + \frac{L^2 }{r^2}
\right]  \tilde{\Phi}(r) = 0\,,
\label{eqbetaadsn}
\end{align}
where $L^2$ is the separation constant. Solving Eq.\ \eqref{eqbetaadsn} around spatial infinity,
we obtain the asymptotic behavior of the solution as
\begin{align}
\tilde{\Phi}(r) = r^{\Delta_+} 
\sum_{i=0}\frac{c_{+}^{(i)}}{r^{2i}} +  r^{-\Delta_-}\sum_{i=0}\frac{c_{-}^{(i)}}{r^{2i}}\,,
\end{align}
where $\Delta_\pm =  \pm(1-n \pm \sqrt{(n-1)^2 + 4m^2})/2$ and 
$c^{(i)}_\pm = (-1)^iL^{2i} c_\pm^{(0)} 
\prod_{j=1}^{i} ((\pm \Delta_\pm + n - 2j-1)(\pm \Delta_\pm - 2j) - m^2)^{-1}$ for $i\ge1$
and $c_\pm^{(0)}$ are constant. 
The two modes with the leading terms $r^{\Delta_+}, r^{-\Delta_-}$
are called non-normalizable and normalizable modes, respectively.
{}From \eqref{eqdefkpm}, we can see $\Delta_\pm = \pm k_\pm$ for the above ladder operators.
Acting the ladder operators on $\Phi$, a non-normalizable mode is mapped into a non-normalizable mode, and 
a normalizable mode is mapped into a normalizable mode unless 
$m^2$ or $m^2 + \delta m^2$ is between 
$m_{BF}^2$ and $m_{BF}^2 + 1$, where $m_{BF}^2 := -(n-1)^2/4$ is the BF bound mass.
If the mass is between the above region then two modes are normalizable. 
Note that the ladder operators do not necessarily keep
the form of separation of variables due to the derivative with respect to $x^A$.

In particular, $D_{a_1,-k-n+2} D_{a_2,k}$ maps a solution of KGE to 
another solution of the same KGE, we can obtain variety of solutions from a single seed solution.
Since the ladder operators are regular everywhere, 
if a seed solution $\Phi$ is regular, $D_{a_1,-k-n+2} D_{a_2,k}\Phi$ is also regular. From the 
point of view of AdS/CFT correspondence  the ratio of the coefficients between 
normalizable and non-normalizable modes is the expectation value of the operator.
If the asymptotic behavior of $D_{a_1,-k-n+2} D_{a_2,k} \Phi$ is different from $\Phi$,
this corresponds to different physical situation.
If we use $D_{-1}$, we can show $D_{-1,-k-n+2} D_{-1,k}\Phi = - L^2 \Phi$ for a solution 
with the separation of variables form $\Phi = \alpha(x^A) \tilde{\Phi}(r)$.
If we use other ladder operators, $D_{a_1,-k-n+2} D_{a_2,k}\Phi$ is different from $\Phi$.

We comment on massless scalar fields in $AdS_5 \times S^5$.
The massless KGE in $AdS_5 \times S^5$ reduces to the effective massive KGE in $AdS_5$
\begin{align}
(\square_{AdS_5} - \Lambda \ell(\ell + 4) )\Phi = 0,
\label{effectivekgeAdS5S5}
\end{align}
where $\ell$ denotes the different Kaluza-Klein modes.
The mass spectrum corresponds to the masses which can be mapped from massless 
scalar fields in $AdS_5$ by using the mass ladder operators.
This implies that
there is a duality among the zero mode and 
Kaluza-Klein modes on massless scalar fields in $AdS_5 \times S^5$.

\section{Ladder operators in sphere and spherical harmonics}
By applying our formulation to the 2-dimensional sphere $S^2$, we obtain three ladder operators
\begin{align}
D_{1,k} &= \cos\theta \cos \phi \frac{\partial}{\partial \theta} 
- \frac{\sin \phi}{\sin \theta} \frac{\partial}{\partial \phi} + k \sin\theta \cos\phi
\\
D_{-1,k} &=  \cos \theta\sin \phi  \frac{\partial}{\partial \theta} 
+ \frac{\cos \phi}{\sin \theta} \frac{\partial}{\partial \phi} + k \sin\theta \sin\phi
\\
D_{0,k}
&= \sin\theta \frac{\partial}{\partial\theta} - k \cos\theta\,,
\end{align}
where $\theta$ and $\phi$ are spherical coordinates on $S^2$ in which the metric is given by $ds^2_{S^2} = d\theta^2 + \sin^2\theta d\phi^2$.
The ladder operators map solutions of the eigenvalue equation for the Laplacian $\Delta_{S^2}$ with eigenvalues $\lambda =-k(k+1)$,
\begin{equation}
(\Delta_{S^2} - \lambda )\Phi = 0\,,
\label{helmholtzeq}
\end{equation}
into solutions with eigenvalues $\lambda =-k(k-1)$. It should be noted that $k$ is not necessarily integer. If $k$ is an integer $\ell$, spherical harmonics\footnote{
The normalized spherical harmonics are given by $\sqrt{(2\ell +1)/4\pi}\sqrt{ (\ell-m)!/(\ell+m)!} Y_{\ell,m}$.
} $Y_{\ell,m}= P_\ell^m(\cos\theta)e^{im\phi}$ 
are the eigenfunctions for $\Delta_{S^2}$ with $\lambda = - \ell(\ell+1)$,
and the ladder operators change the quantum number $\ell$, while the usual ladder operators $L_\pm$ constructed from the spherical symmetry change the quantum number $m$. Introducing $D_{\pm,k} = D_{1,k} \pm i D_{-1,k}$, we can reproduce the relations in \cite{Higuchi:1986wu,Gelderen:1998}
\begin{eqnarray}
 D_{+,\ell} Y_{\ell,m}
&=& Y_{\ell-1,m+1} \,, \label{rec_rel_11} \\
 D_{-,\ell} Y_{\ell,m}
&=& -(\ell+m)(\ell+m-1)Y_{\ell-1,m-1} \,, \label{rec_rel_12} \\
 D_{0,\ell} Y_{\ell,m}
&=& - (\ell+m) Y_{\ell-1,m} \,, \label{rec_rel_13} 
\end{eqnarray}
and
\begin{eqnarray}
 D_{+,-\ell} Y_{\ell-1,m}
&=& Y_{\ell,m+1} \,, \label{rec_rel_21} \\
 D_{-,-\ell} Y_{\ell-1,m}
&=& -(\ell-m)(\ell-m+1) Y_{\ell,m-1} \,, \label{rec_rel_22} \\
 D_{0,-\ell} Y_{\ell-1,m}
&=& (\ell-m) Y_{\ell,m} \,. \label{rec_rel_23} 
\end{eqnarray}
These are useful relations to obtain the entire spectrum of the Laplacian on $S^2$. Their relation to geometry of $S^2$ had never been uncovered; we stress that
the conformal symmetry of $S^2$ is crucial for the existence of such ladder operators. We should note that we can also apply our formalism to higher dimensional spheres $S^n$.

Solutions of \eqref{helmholtzeq} which are not spherical harmonics will have a singular behavior. 
However, it is possible that the ladder operator 
can map such singular solution to a regular one~\footnote{
Here, we call a local solution regular if the domain of the solution can be extended to the whole of $S^2$; otherwise singular, that is, the domain of the solution cannot be extended to the whole of $S^2$.
}. 
For example, if we consider 
$\Phi = e^{i\phi}/\tan\theta $ 
which satisfies $\Delta_{S^2} \Phi = 0$ and is
singular at the pole, we can show $D_{0,-1} \Phi = Y_{11}$.

\section{Relation with supersymmetric quantum mechanics}
\label{susyquantummechanics}
The concept of ladder operators was developed in the context of exactly solvable systems in quantum mechanics. One is thus naturally led to 
inquire whether mass ladder operators can be framed in this context as well.
In fact, one can obtain shape invariant potentials \cite{Gendenshtein:1984vs} in supersymmetric quantum mechanics from the KGE,\footnote{In a series of works~\cite{Evnin:2015gma, Evnin:2015wyi, Evnin:2016nsz}, 
the relation between a quantum mechanics system with a shape invariant potential
and the KGE in $AdS$ spacetime was shown, and the structure of the hidden symmetry of them was also discussed.}
where our ladder operators are regarded as supercharges \cite{Witten:1981nf}.
We make a conformal transformation $\bar{g}_{\mu \nu} = \Omega^2 g_{\mu \nu}$ with an appropriate conformal factor $\Omega$ such that a CCKV $\zeta^\mu$ for $g_{\mu\nu}$ is transformed into a Killing vector for $\bar{g}_{\mu \nu}$. Then, the massive KGE $(\Box-m^2)\Phi = 0$ for $g_{\mu \nu}$ is written in terms of $\bar{g}_{\mu \nu}$ as
\begin{align}
\left[\partial^2_{\bar{\lambda}}
+ \tilde{\square} - V(\bar{\lambda}, m^2) \right] \bar{\Phi}=0\,,
\end{align}
where $\bar{\Phi} = \Omega^{(2-n)/2}\Phi$, $\partial_{\bar{\lambda}} = \zeta^\mu \partial_\mu$ is a Killing vector and $\tilde{\square}$ is the Laplacian (or d'Alembertian) on an $(n-1)$-dimensional space (or spacetime).
Thus, with the separation of variables $\bar{\Phi}=\psi(\bar{\lambda})\Theta(x^i)$, we obtain the Schr\"odinger equation in one dimension,
\begin{align}
 \left[-\partial^2_{\bar{\lambda}} 
 + V(\bar{\lambda}, m^2)\right]\psi = E \psi \,,
\end{align}
where $E$ is the separation constant. The potential $V$ is given, up to a constant, by $1/\cos^2 \bar{\lambda}$, $1/\cosh^2\bar{\lambda}$ or $1/\bar{\lambda^2}$. These are known as shape invariant potentials in supersymmetric quantum mechanics, the mass ladder operators being regarded as supercharges (see Appendix.\ref{conformaltr} for details).

\section{Aretakis constants}
Mass ladder operators also appear naturally in black hole physics.
In Refs.~\cite{Aretakis:2011ha, Aretakis:2011hc, Lucietti:2012xr},
it has been shown that an extreme Reissner-Nordstr\"om black hole is linearly unstable. 
In their analysis, a certain quantity (``Aretakis constant''), conserved only on the horizon, plays an important role.
We now show that such constants can be constructed from our ladder operator, in four-dimensional extreme Reissner-Nordstr\"om black holes 
(more details can be found in Appendix.\ref{aretakisconst}).

The near horizon geometry of extreme Reissner-Nordstr\"om black holes
is described by $AdS_2 \times S^{2}$, and massless scalar fields on this spacetime
behave as a massive scalar field on $AdS_2$ with an effective mass $m^2 = \ell(\ell + 1)$
where $\ell$ is azimuthal quantum number of the spherical harmonics.
Thus, we focus on the KGE on $AdS_2$ with this mass. The metric of $AdS_2$ in ingoing Eddington-Finkelstein coordinate is
\begin{align}
ds^2 = - r^2 dv^2 + 2dv dr.
\end{align}
Take solutions $\Phi$ of the KGE, $(\Box - m^2)\Phi = 0$, with $m^2 = \ell(\ell + 1), (\ell = 0,1, \cdots)$ on this spacetime.
Then, one can show that 
\begin{align} 
\partial_v \partial_r^{\ell+1}\Phi |_{r=0} = 0.
\end{align}
Thus, the quantities $\partial_r^{\ell+1}\Phi|_{r=0}$ are constant on the 
Poincar\'e horizon $r=0$. This is the Aretakis constant in $AdS_2$ \cite{Lucietti:2012xr}.

While the quantity $\partial_r^{\ell+1}\Phi$ is not constant outside of $r = 0$, 
since $AdS_2$ is maximally symmetric, we may expect the existence of quantities 
which are constants on every out-going null hypersurface. In fact, we can show
\begin{equation}
 \left(\partial_v + \frac{r^2}{2}\partial_r\right)
 \left[\left(\frac{vr}{2}+1\right)^{2(\ell+1)}\partial_r^{\ell+1} \Phi\right]
 = 0 \,. \label{Aretakis_2}
\end{equation}
Since $\partial_v + (r^2/2) \partial_r$ is an out-going null vector field, 
\begin{align}
A_\ell \equiv \left(\frac{vr}{2}+1\right)^{2(\ell+1)}\partial_r^{\ell+1} \Phi,
\end{align}
is indeed constant on every out-going null hypersurface, and 
$A_\ell$ coincides with the Aretakis constant on $r = 0$. 
In this sense, $A_\ell$ is a generalization of the Aretakis constant.

In $AdS_2$, the operator $D_k$ changes the mass squared from $k(k+1)$ into $(k-1)k$.
So, $D_1 D_2 \cdots D_{\ell-1} D_{\ell}$ maps a massive scalar field with $m^2 = \ell (\ell +1)$
into massless scalar field.
Since we can solve the two dimensional massless KGE, 
we can write 
\begin{align}
D_1 D_2 \cdots D_{\ell-1} D_{\ell}\Phi = F(x^+) + G(x^-),
\end{align}
where
we used the double null coordinates $(x^+, x^-)$. 
Thus $\partial_{x^-}D_1 D_2 \cdots D_{\ell-1} D_{\ell}\Phi = G^\prime(x^-)$ is constant on 
every outgoing null hypersurface $x^- = {\rm const.}$ 
In fact this coincides with $A_\ell$ up to a function of $x^-$. 
Note that the choice of CCKVs $\zeta_{-1}, \zeta_0, \zeta_1$ does not affect this conclusion.
If we consider Reissner-Nordstr\"om black hole spacetime without taking near horizon limit,
we can still derive the Aretakis constant on the horizon in a similar way.
Since there is a relation between Aretakis constant and Newman-Penrose constant~\cite{Bizon:2012we},
the present analysis suggests that Newman-Penrose constant also can be
constructed from our ladder operator.

\section{Discussion}

We developed a mass ladder operator formalism for the massive KGE
and explicitly constructed the operators for $AdS_n$ and $S^2$.
It is possible, and we showed that this happens on $S^2$, that the ladder operator maps a singular to a regular solution
even if CCKVs and the associated functions are regular.
Naturally, in the context of AdS/CFT correspondence regular solutions are preferred objects. 
However, the property above might help in providing a physical interpretation to singular solutions.

The ladder operators on $S^2$ were originally obtained by embedding $S^2$ into three dimensional Euclid space $E^3$ \cite{Gelderen:1998} or sphere $S^3$ \cite{Higuchi:1986wu}.
The harmonic functions on $E^3$ are known as regular and irregular solid harmonics. According to \cite{Gelderen:1998}, taking the covariant derivatives of the solid harmonics along $
\partial_x$, $\partial_y$ and $\partial_z$, yields differential recurrence relations between the solid harmonics with different azimuthal and magnetic quantum numbers. By restricting the recurrence relations onto $S^2$, we obtain the ladder relations for the spherical harmonics.
Higuchi \cite{Higuchi:1986wu} also constructed the symmetric tensor harmonics on $S^n$ in the reductive construction, where $S^n$ is embedded into $S^{n+1}$. This suggests the existence of the ladder operators for vector or tensor fields on $S^n$ and also maximally symmetric spacetimes.

Another interesting direction is to consider higher-order operators. For symmetry of the Laplace equation or KGE in a curved spacetime, they have been studied by many authors \cite{Boyer:1976,Carter:1977pq,Eastwood:2002su,Visinescu:2010zzb}. While our formulation in this paper focused on first-order mass ladder operators, it would be of great interest to consider higher-order mass ladder operators if there exists a curved spacetime which admits a crucial higher-order operator not reducible to first-order operators.

We also showed the relation between these operators and supersymmetric quantum mechanics potentials having
shift shape invariance. If we start from generic 1-dimensional supersymmetric quantum mechanics potential, 
we can expect to obtain a class of scalar fields with potential which has a ladder structure.

As an application, we constructed Aretakis constant from mass ladder operators on $AdS_2$.
If we consider Reissner-Nordstr\"om spacetimes without taking the near horizon limit,
the Aretakis constant on the horizon can be derived in a similar way.
This suggests the intriguing possibility of mass ladder operators being useful constructs also for less symmetric spacetimes, with only approximate conformal symmetry.

In Minkowski spacetime, the existence of mass ladder operators 
(shown in Appendix~\ref{anotherladder}) is not surprising,
as there is no scale in the problem other than the mass parameter in the massive KGE.
In curved spacetimes however, the different hierarchy (as compared to curvature scale) in the mass of scalar fields
is expected to play a fundamental role.
Notwithstanding, if we consider the curved spacetimes which admit mass ladder operators 
(including the maximally symmetric spacetimes), 
solutions of KGE between different masses are connected.
Furthermore, the map induced by the ladder operator is surjective (or onto),
so all the solutions with mass squared $m^2 + \delta m^2$ can be constructed 
from the solutions with mass squared $m^2$.
This suggests that the physical properties of KGE with different masses,
which are connected by the ladder operator, are very similar contrary to the naive expectation.

\section*{Acknowledgments}
The authors would like to thank 
Oleg Evnin,
Shunichiro Kinoshita, Masato Minamitsuji, Keiju Murata, Harvey S. Reall, Kentaro Tanabe, Norihiro Tanahashi and Benson Way for their useful comments. 
V.C. and M.K. acknowledge financial support provided under the European Union's H2020 ERC 
Consolidator Grant ``Matter and strong-field gravity: New frontiers in Einstein's theory''
grant agreement no. MaGRaTh-646597, and under the H2020-MSCA-RISE-2015 Grant No. StronGrHEP-690904.
T.H. was supported by the JSPS Grant-in-Aid for Scientific Research JP14J01237 in the early stage of this project, and was supported in part by the Supporting Program for Interaction-based Initiative Team Studies (SPIRITS) from Kyoto University and by a JSPS Grant-in-Aid for Scientific Research (C) No.\,15K05051.
M.K. was partially supported by a grant for research abroad from JSPS.
M.K. would like to thank the conference ``Gravity - New perspectives from strings and higher dimensions 2015'', where this work was initiated, and Theoretical Astrophysics Group and 
Yukawa Institute for Theoretical Physics at Kyoto University and
Theoretical Astrophysics Group at Osaka City University for their hospitality.

\appendix

\section{Explicit metric form admitting a mass ladder operator}
\label{explicitmetriccckv}
In \cite{Batista:2014uja}, 
all canonical forms for metrics admitting a CCKV, denoted by $\zeta^\mu$, were investigated in arbitrary dimensions. In Lorentzian signature, they are classified 
according to whether $\zeta^\mu$ is null or not. In the null case, $\zeta^\mu$ becomes a covariantly constant null vector, so that $Q=0$ and $R^\mu{}_\nu\zeta^\nu = 0$. Hence, the operator does not become a ladder operator for the d'Alembertian. In the non-null case, it is possible to introduce a function $\lambda$, called the potential of $\zeta^\mu$, such that $d\lambda$ is the 1-form dual to $\zeta^\mu$. Using the potential as a coordinate, we can choose a local coordinate system $(x^\mu)=(\lambda,x^i)$. Then, a metric in the case in $n$ dimensions is written as
\begin{equation}
 ds^2 = g_{\mu\nu}dx^\mu dx^\nu
 = \frac{1}{f(\lambda)}d\lambda^2 + f(\lambda)\tilde{g}_{ij}(x)dx^idx^j \,, \label{met_CCKV}
\end{equation}
where $f(\lambda)$ is an arbitrary function and $\tilde{g}_{ij}$ is an $(n-1)$-dimensional metric.
\footnote{If $\zeta^\mu$ is timelike, {\it i.e.}, $f < 0$, 
$-\tilde{g}_{ij}$ should be a positive definite metric so that the metric 
$g_{\mu \nu}$ has $[-,+,+,\cdots,+]$ signature.} 
This metric admits a CCKV\footnote{
It is possible to show that, in addition to (\ref{general_CCKV}), the metric (\ref{met_CCKV}) can admit a CCKV if $\tilde{g}_{ij}$ admits a CCKV. Actually, the CCKV equation $\nabla_\mu\zeta_\nu = Qg_{\mu\nu}$ for the metric (\ref{met_CCKV}) can be solved by
\begin{equation}
 \zeta = -\frac{\tilde{Q}}{2\tilde{\chi}} f^\prime(\lambda) \sqrt{f(\lambda)} \frac{\partial}{\partial\lambda}
 + \frac{1}{\sqrt{f(\lambda)}} \tilde{\zeta}^i \frac{\partial}{\partial x^i} \,, \label{CCKV_Sd}
\end{equation}
where $\tilde{\zeta}^i$ is a CCKV for $\tilde{g}_{ij}$, $\tilde{\nabla}_i\tilde{\zeta}_j = \tilde{Q}\tilde{g}_{ij}$.
The associated function $Q$ of $\zeta^\mu$ is given by
 $Q= 
({\chi}/{\tilde{\chi}}) \sqrt{f(\lambda)} \tilde{Q}$.
}
\begin{equation}
\zeta = f(\lambda) \frac{\partial}{\partial\lambda} \,. \label{general_CCKV}
\end{equation}

If we impose the condition \eqref{Ricci_CCKV} for this spacetime, $f$ takes the form
\begin{equation}
 f(\lambda) = - \chi \lambda^2 + c_1 \lambda + c_0 \,, \label{func_f}
\end{equation}
where $c_0$ and $c_1$ are constant.
With $f(\lambda)$ given by (\ref{func_f}), the ladder operator is
\begin{equation}
 D_k = f(\lambda) \partial_\lambda - \frac{k}{2}f^\prime(\lambda) \,, \label{LadOp_General}
\end{equation}

\section{Surjectivity and kernel}
\label{ontomap}
We can show that $D_k$ is a surjective (onto) map, i.e.,
for arbitrary solution of $(\square - m^2 - \delta m^2)\bar{\Phi} = 0$,
we can find a solution of the equations 
\begin{align}
D_k\Phi &= \bar{\Phi} 
\label{eqI1}
\\  
(\square - m^2 )\Phi &= 0,
\label{eqI2}
\end{align}
where $m^2$ and $\delta m^2$ are given by $m^2 = - \chi k(k+n-1), m^2 + \delta m^2 = -\chi (k-1)(k+n-2)$.
The general solution of \eqref{eqI1} is 
\begin{align}
\Phi = f^{k/2} \left(
\int d\lambda f^{-1 - k/2} \bar{\Phi}  + P(x^i)
\right),
\end{align}
where $P(x^i)$ is arbitrary function of $x^i$.
After a straightforward calculation, we obtain
\begin{align}
& (\square - m^2 )\Phi 
\nonumber \\ &~~ = f^{-1+k/2}\left[
\tilde{\Box} + \frac{k(k+n-2)}{4} (c_1^2 + 4 c_0\chi)
\right] P(x^i),
\nonumber
\end{align}
where we used Eq.\eqref{func_f}.
For $P = 0$ we recover Eq.\eqref{eqI2}, showing that $D_k$ is a surjective map.

If there exists a non trivial solution of the equation $\left[
\tilde{\Box} + k(k+n-2)(c_1^2 + 4 c_0\chi)/4
\right] P(x^i) = 0$, such functional degrees of freedom correspond to the kernel of $D_k$, i.e., 
the solutions of both $D_k \Phi = 0$ and $(\square - m^2)\Phi = 0$.
In particular, if $c_1 = c_0 = 0$, $P = {\rm const}$ is a nontrivial solution, then 
$\Phi = C f^{k/2}$ becomes a kernel of $D_k$.

\section{Another ladder operator for $\chi = 0$, $Q = {\rm const.}$ case}
\label{anotherladder}
The operator $D_k$ relates scalars of different mass if the eigenvalue of the Ricci tensor $\chi$ is not zero.
However, for constant $Q$ a ladder operator can be defined  even for $\chi = 0$ case, albeit in a modified way.
If $Q = c = {\rm const.}$, the conformal Killing equation (4) 
becomes 
the homothetic Killing equation
\begin{equation}
\nabla_\mu\zeta_\nu+\nabla_\nu\zeta_\mu = 2c g_{\mu\nu}.
\label{HomotheticKVequation}
\end{equation}
The commutation relation \eqref{massladdercommutator1} 
with $k = 0$ is
\begin{equation}
    [\Box, {\cal L}_\zeta ] = \chi(n-2) {\cal L}_\zeta  + 2c \Box.
\end{equation}
If $\chi$ is zero, we can define another ladder operator 
$\tilde{D}_\lambda := e^{\lambda {\cal L}_\zeta} = \sum_{j=0}^\infty (j!)^{-1}(\lambda {\cal L}_\zeta)^j$ 
with a parameter $\lambda$, which satisfies the commutation relation\footnote{ We can show this relation by using the equation
$ [\Box, ({\cal L}_\zeta)^n ] = ((2c + {\cal L}_\zeta)^n - ({\cal L}_\zeta)^n )\Box, $
where $n$ is a positive integer.}
\begin{align}
[\Box,\tilde{D}_\lambda ] = (e^{2 \lambda c} - 1)\tilde{D}_\lambda \Box.
\end{align}
Acting on a scalar field $\Phi$, we obtain
\begin{align}
\Box \tilde{D}_\lambda \Phi - e^{2 \lambda c} \tilde{D}_\lambda  \Box\Phi = 0.
\label{eqh3}
\end{align}
If $\Phi$ satisfies a massive KGE, then \eqref{eqh3} becomes
\begin{align}
(\Box - e^{2 \lambda c}m^2 )\tilde{D}_\lambda \Phi = 0.
\end{align}
This shows that $\tilde{D}_\lambda$ maps a scalar field with $m^2$ to that with $e^{2 \lambda c}m^2$.
Since the parameter $\lambda$ is an arbitrary number, $\tilde{D}_\lambda$ can change the mass continuously.
Note that $\tilde{D}_\lambda$ cannot change the signature of the mass squared, but can change 
the absolute value. In Minkowski spacetime $g_{\mu \nu} = \eta_{\mu \nu}$, 
we can explicitly construct the ladder operator as
$\tilde{D}_\lambda = e^{\lambda (x^\mu \partial_\mu + \xi^\mu \partial_\mu)}$, 
where $\xi^\mu$ is an arbitrary Killing vector on $\eta_{\mu \nu}$.

\section{Regularity of ladder operators}
\label{ladderregularityadsn}
To see the regularity of the ladder operator on $AdS_n$ 
beyond the Poincar\'e horizon, introduce global coordinates 
\begin{align}
r &= \frac{\cos \tau - \Omega_{n-1} \sin \rho}{\cos \rho},
\\
t &=\frac{ \sin \tau}{\cos \tau - \Omega_{n-1} \sin \rho},
\\
x^i &= \frac{\Omega_i \sin \rho}{\cos \tau - \Omega_{n-1} \sin \rho}, ~~~~(i = 1,2, \cdots,n-2),
\end{align}
where $\Omega_{i}$ satisfy the relation $\sum_{i=1}^{n-1} \Omega_i^2 = 1$.
In this coordinate system, the metric becomes
\begin{align}
ds^2 = \frac{1}{\cos^2\rho}(-d\tau^2 + d\rho^2 + \sin^2\rho \sum_{i=1}^{n-1}d\Omega_i^2).
\end{align}
Spatial infinity corresponds to $\rho = \pm \pi/2$.
Note that $\sum_{i=1}^{n-1}d\Omega_i^2$ is the metric of a $(n-2)$-dimensional unit sphere.
The associated functions of CCKVs $Q_a, (a=-1,0,1,.\cdots,n-1)$ in these coordinates are 
\begin{align}
Q_{-1} &= \frac{\cos \tau - \Omega_{n-1} \sin \rho}{\cos \rho},
\\
Q_0 &= \frac{ \sin \tau}{\cos \rho},
\\
Q_{i} &=  \frac{\Omega_i \sin \rho}{\cos \rho}, ~~(i=1,2,\cdots,n-2)
\\
Q_{n-1} &= \frac{\cos \tau +\Omega_{n-1} \sin \rho  }{\cos\rho}.
\end{align}
Thus, $Q_a$ is finite except at spatial infinity.
Since the 1-form $d\Omega_{n-1}$ is regular (except at the sphere's pole),
the 1-forms $dQ_a$ are also regular in $-\pi/2 < \rho < \pi/2$.
In $AdS_n$, $dQ_a = \zeta_{a,\mu}dx^\mu$, 
so CCKVs $\zeta_a^\mu$ and the ladder operators $D_{a,k}$ are regular in $-\pi/2 < \rho < \pi/2$.

\section{Conformal transformation and supersymmetric quantum mechanics}
\label{conformaltr}
Given a CKV $\zeta^\mu$ for a metric $g_{\mu\nu}$, we can make a conformal transformation $\bar{g}_{\mu\nu}=\Omega^2g_{\mu\nu}$ under which $\zeta^\mu$ is  a Killing vector. We have already seen that if a spacetime admits a CCKV $\zeta^\mu$, the metric and CCKV have the forms \eqref{met_CCKV} and \eqref{general_CCKV}, respectively. Hence, by setting $\Omega = 1/\sqrt{f}$, the CCKV $\zeta^\mu$ for $g_{\mu\nu}$ becomes a Killing vector for $\bar{g}_{\mu \nu}$. Under this conformal transformation, we have
\begin{align}
 (\square - m^2)\Phi
 = \Omega^{(n+2)/2}\Big(\bar{\square} 
  - V(\lambda, m^2) \Big) \bar{\Phi}\,,
\end{align}
where $\bar{\Phi} = \Omega^{(2-n)/2}\Phi$ and
\begin{align}
 V(\lambda, m^2) =(16 m^2 f + (n-2)^2 (f^\prime)^2
+ 4(n-2)f f^{\prime \prime})
/16 \,.
\end{align}
Hence, massive KGE on $g_{\mu \nu}$, $(\square - m^2)\Phi = 0$ leads to 
\begin{align}
\bar{\square} \bar{\Phi} - V(\lambda, m^2) \bar{\Phi} = 0 \,,
\label{eqkgeqconformaltr}
\end{align}
where $\bar{\square}$ is the d'Alembertian on $\bar{g}_{\mu \nu}$.
In addition, if we assume the function $f(\lambda)$ is given by \eqref{func_f}, the potential $V$ becomes a quadratic polynomial of $\lambda$,
\begin{align}
V &= s_0 + s_1 \lambda + s_2 \lambda^2
\end{align}
with the coefficients
\begin{align}
s_0 &= c_1^2 (n-2)^2/16  +  c_0 (m^2 + \chi (1 - n/2)),
\\
s_1 &= c_1(m^2 - (n-2)n \chi/4),
\\
s_2 &= \chi (-4m^2 + (n-2)n\chi)/4.
\end{align}
Furthermore, we introduce the coordinate $\bar{\lambda}$ as $\partial_{\bar{\lambda}}=\zeta^\mu\partial_\mu=f\partial_\lambda$. Since $\bar{\square} \bar{\Phi} = f \partial_\lambda \left[f \partial_\lambda \bar{\Phi}\right]
+\tilde{\square} \bar{\Phi}$, \eqref{eqkgeqconformaltr} is
\begin{align}
&\frac{\partial^2}{\partial \bar{\lambda}^2} \bar{\Phi}+ 
\tilde{\square} \bar{\Phi}
+\frac{(c_1^2 + 4 c_0 \chi)}{16 \chi}  \nonumber\\
& \times \Bigg[\frac{4 m^2 - n(n-2)\chi}{\cos^{2}(\bar{\lambda}\sqrt{-c_1^2 - 4 c_0 \chi}/2)}+(n-2)^2\chi
\Bigg]\bar{\Phi}=0
\end{align}
where $\tilde{\square}$ is the d'Alembertian on an $(n-1)$-dimensional spacetime. Imposing $[\tilde{\square}, \partial_{\bar {\lambda}}] = 0$ and $[\tilde{\square}, Q] =0$, the separation of variables $\bar{\Phi}=\psi(\bar{\lambda})\Theta(x^i)$ leads to the Schr\"odinger equation in one dimension
\begin{align}
 H(m^2)\psi \equiv \left[-\frac{d^2}{dz^2} + V(m^2,z)\right] \psi = E \psi \,,
\end{align}
where we have introduced the coordinate $z=\bar{\lambda}\sqrt{-(c_1^2+4c_0\chi)}/2$, in which the potential is given by
\begin{align}
 V(m^2,z) = \frac{m^2/\chi-n(n-2)/4}{\cos^2z} + \frac{(n-2)^2}{4} \,,
\end{align}
and $E$ is a separation constant. This is known as a shape invariant potential in supersymmetric quantum mechanics. This form of potential changes to $1/\cosh^2z$ if the signature of $-c_1^2 - 4 c_0 \chi$ is negative. We have assumed that either $c_0$ or $c_1$ are non-vanishing. When $c_0=c_1=0$, the potential becomes 
a quadratic polynomial of
$1/\bar{\lambda}$, as in the problem of the hydrogen atom.

The present conformal transformation transforms the ladder operator $D_k$ for $\square$ into the ladder operator $\bar{D}_k = \Omega^{(2-n)/2}D_k\Omega^{-(2-n)/2}$ for $\bar{\square}$. To be explicit, it is written in the coordinate $z$ as
\begin{align}
 \bar{D}_k = \frac{d}{dz}
 - \left(k-\frac{2-n}{2}\right) \tan z + \textrm{const.}
\end{align}
In supersymmetric quantum mechanics, the Hamiltonian $H$ and supercharge $Q$ are related via $Q^2=H$. In the present case, this can be realized by setting
\begin{align}
 H =& \left(
\begin{array}{cc}
H(m^2) &0  \\
0 &H(m^2+\delta m^2)
\end{array}
\right) \,, \\
Q =& \left(
\begin{array}{cc}
0 &-\bar{D}_{-k-n+2} \\
\bar{D}_k &0 
\end{array}
\right) \,.
\end{align}
Thus, our ladder operator corresponds to the supercharge in supersymmetric quantum mechanics.

\section{Aretakis constants}
\label{aretakisconst}
\subsection{Mass ladder operators in $AdS_2$}
In double null coordinates, $AdS_2$ metric is given by
\begin{equation}
ds^2_{AdS_2} = - \frac{4|\Lambda|}{(x^+-x^-)^2}dx^+ dx^- \,, \label{met_ads2}
\end{equation}
where $1/\sqrt{|\Lambda|}$ is the AdS radius. Setting $\Lambda=1$, the KGE \eqref{eq:massivescalar}
is given by
\begin{equation}
-(x^+-x^-)^2\partial_+ \partial_- \Phi = m^2 \Phi \,. \label{KG_equation}
\end{equation}
There are an infinite number of CKVs on $AdS_2$, described by two copies of the Witt algebras. Since the Witt algebra contains $SO(2,1)$ subalgebra, there are six CKVs as generators for the $SO(2,2)=SO(2,1)\times SO(2,1)$ subalgebra. Three of them are KVs, 
\begin{eqnarray}
\xi_{-1} &=& \partial_+ + \partial_- \,,\\
\xi_0  &=& x^+\partial_+ + x^-\partial_- \,, \\
\xi_1  &=& (x^+)^2\partial_+ + (x^-)^2\partial_- \,, \label{kv}
\end{eqnarray}
and the other ones are CCKVs, which are given by
\begin{eqnarray}
\zeta_{-1}  &=& \partial_+ - \partial_- \,, \\
\zeta_0  &=& x^+ \partial_+ - x^- \partial_- \,, \\
\zeta_1  &=& (x^+)^2 \partial_+ - (x^-)^2 \partial_- \,. \label{cckv}
\end{eqnarray}
Since $AdS_2$ admits three CCKVs, we are able to construct three one-parameter families of mass ladder operators
\begin{eqnarray}
&& D_{-1,k} = \partial_+ - \partial_- + \frac{2k}{x^+-x^-} \,, \label{op1} \\
&& D_{0,k} = x^+ \partial_+ - x^- \partial_- + \frac{k(x^++x^-)}{x^+-x^-} \,, \label{op2} \\
&& D_{1,k} = (x^+)^2 \partial_+ - (x^-)^2 \partial_- 
+ \frac{2kx^+x^-}{x^+-x^-} \,, \label{op3} 
\end{eqnarray}
where $k$ is a real parameter.
$D_{i,k}$ map a solution to the KGE on $AdS_2$ with mass squared $k(k+1)$ into another solution with mass squared  $k(k-1)$.

It should be emphasized that for any solution satisfying the BF bound, $m^2\geq -1/4$, two operators $D_{i,k_\pm}$ exist for each $i=-1,0,1$. For a fixed $m^2$, the corresponding two values for $k$ are given by
\begin{equation}
k_\pm = \frac{-1\pm\sqrt{1+4m^2}}{2} \,.
\end{equation}
Especially in the range $0\leq m^2$, one of the two operators is a mass raising and the other a mass lowering operator. If $-1/4\leq m^2 <0$, both become mass raising operators.

If $k$ is a natural number, then $m^2$ is shifted by $D_{i,-k}$ and $D_{i,k}$ as follows:
\begin{widetext}
\begin{equation}
\cdots
\underset{D_{i,-(k+1)}}{\overset{D_{i,k+1}}{\quad\rightleftharpoons\quad}}
k(k+1) 
\underset{D_{i,-k}}{\overset{D_{i,k}}{\quad\rightleftharpoons\quad}} 
(k-1)k 
\underset{D_{i,-(k-1)}}{\overset{D_{i,k-1}}{\quad\rightleftharpoons\quad}} 
\cdots
\underset{D_{i,-3}}{\overset{D_{i,3}}{\quad\rightleftharpoons\quad}} 6
\underset{D_{i,-2}}{\overset{D_{i,2}}{\quad\rightleftharpoons\quad}} 2
\underset{D_{i,-1}}{\overset{D_{i,1}}{\quad\rightleftharpoons\quad}} 0
\end{equation}
\end{widetext}
By acting the mass lowering operators repeatedly on a massive scalar field of $m^2 =k(k+1)$, we can annihilate the mass. Hence, we obtain the operator
\begin{equation}
D^{(k)}_{i_1,i_2,\cdots, i_k} = D_{i_k,1}\cdots D_{i_2,k-1}D_{i_1,k} \,,
\label{completemassannihilation}
\end{equation}
which map a scalar field with mass $k(k+1)$ into a massless scalar field. 
By using $D^{(\ell)}_{i_1,i_2,\cdots, i_\ell}$ in Eq.\eqref{completemassannihilation},
we can construct conserved quantities on every out-going null hypersurface.
Since $D^{(\ell)}_{i_1,i_2,\cdots, i_\ell} \Phi_\ell$ satisfies a two dimensional massless KGE,
we can write
$D^{(\ell)}_{i_1,i_2,\cdots, i_\ell} \Phi_\ell = \phi_+(x^+) + \phi_-(x^-)$, where
 $\phi_\pm(x^\pm)$ are arbitrary functions of $x^\pm$, respectively.
Taking the derivative w.r.t. $x^-$ of $D^{(\ell)}_{i_1,i_2,\cdots, i_\ell} \Phi_\ell $,
the quantity
\begin{align}
\partial_{-}D^{(\ell)}_{i_1,i_2,\cdots, i_\ell} \Phi_\ell = \partial_- \phi_-(x^-)
\end{align}
is constant on out-going null hypersurfaces $x^+ = {\rm const.}$

\subsection{Aretakis constants in $AdS_2$}
In ingoing Eddington-Finkelstein coordinates, $AdS_2$ metric is written in the form
\begin{equation}
 ds^2 = - r^2 dv^2 + 2dv dr \,, \label{met_ads2_2}
\end{equation}
where the $AdS$ radius has been already taken to be unit.
The Poincar\'e horizon is located at $r=0$, which is an out-going null hypersurface. In the present coordinates, the KGE with $m^2 = \ell(\ell + 1), (\ell = 0,1, \cdots)$ is given by
\begin{equation}
2\partial_v\partial_r\Phi_\ell + \partial_r(r^2\partial_r\Phi_\ell) - \ell(\ell + 1) \Phi_\ell = 0\,.
\end{equation}
From this equation, it follows that
\begin{equation}
 \partial_v \partial_r^{\ell+1}\Phi_k \Big|_{r=0} = 0 \,,
 \label{Aretakis_equation}
\end{equation}
where $\Phi_\ell$ is a solution for $m^2 =\ell(\ell+1)$. The quantities $H_\ell\equiv \partial_r^{\ell+1}\Phi_\ell|_{r=0}$ are known as Aretakis constants~\cite{Lucietti:2012xr}. 
The quantities $\partial_r^{\ell+1}\Phi_\ell$ are constants 
on the Poincar\'e horizon, but not outside. However, since $AdS_2$ is maximally symmetric, we expect the existence of quantities which are constants on every out-going null hypersurface. In fact, (\ref{Aretakis_equation}) can extend to the outside of the Poincar\'e horizon, and we obtain
\begin{equation}
 \left(\partial_v + \frac{r^2}{2}\partial_r\right)
 \left[\left(\frac{vr}{2}+1\right)^{2(\ell+1)}\partial_r^{\ell+1} \Phi_\ell\right]
 = 0 \,. \label{Aretakis_3}
\end{equation}
Hence, we define the quantity
\begin{equation}
A_\ell \equiv \left(\frac{vr}{2}+1\right)^{2(\ell+1)}\partial_r^{\ell+1} \Phi_\ell \,, 
\label{Aretakis_4}
\end{equation}
which coincides with the Aretakis constant $H_\ell$ at the Poincar\'e horizon. Since $\partial_v + (r^2/2) \partial_r$ is an out-going null vector field, $A_\ell$ is indeed constant on every out-going null hypersurface. In what follows, we still call these Aretakis constants.
For the metric form (\ref{met_ads2_2}), we arrange the coordinate transformation $x^+=v$ and $x^-=v + 2/r$ and obtain the metric form (\ref{met_ads2}). Since we have $\partial_+ = \partial_v +(r^2/2)\partial_r$ and $\partial_- = -(r^2/2)\partial_r$, (\ref{Aretakis_3}) is
\begin{equation}
 \partial_+ A_\ell = 0 \,, \label{sadfas}
\end{equation}
which means that $A_\ell$ is a solution to the massless KGE.

In double null coordinates, Eq.\ (\ref{Aretakis_4}) is written as
\begin{equation}
 A_\ell = (x^-)^{2(\ell+1)} L_-^{(\ell+1)} \Phi_\ell ,
\end{equation}
where
\begin{eqnarray}
 L_-^{(\ell+1)} \equiv \frac{1}{(x^+-x^-)^{2(\ell+1)}}
 \left[(x^+-x^-)^2\partial_- \right]^{\ell+1} . \label{Lop1}
\end{eqnarray}
Since there is symmetry between $x^+$ and $x^-$, we can also define 
an operator  $L_+^{(\ell+1)}$
\begin{eqnarray}
 L_+^{(\ell+1)} \equiv \frac{1}{(x^+-x^-)^{2(\ell+1)}}
 \left[(x^+-x^-)^2\partial_+ \right]^{\ell+1}, \label{Lop}
\end{eqnarray}
which can give a conserved quantity $L_+^{(\ell+1)} \Phi_\ell$ on every ingoing null hypersurface.\footnote{The ladder operators $L^{(k)}_\pm$ can be written in the covariant form
\begin{equation}
 L_\pm^{(\ell)} = K_{(\pm)}^{\mu_1\mu_2\dots \mu_k}\nabla_{\mu_1}
 \nabla_{\nu_2}\cdots\nabla_{\mu_\ell} \,,
\end{equation}
where $K_{(\pm)}^{\mu_1\mu_2\cdots \mu_\ell}$ are conformal Killing-St\"akel tensors. In the double null coordinates $(x^+,x^-)$, the nonzero components are given by $K_{(+)}^{++\cdots +}=1$ and $K_{(-)}^{--\cdots -}=1$. Although this fact might be suggesting that our construction of the ladder operators can be extended to a wider framework in which higher-rank conformal Killing-St\"akel tensors play an important role, we leave it as a future problem.}

We explicitly checked this quantity is
equal to $A_\ell$ up to some function of $x^-$ for $\ell = 0, 1, 2$.
We should note that regardless of the choice of $\zeta_i$, this quantity provides an Aretakis constant.
We conjecture that $\partial_{-}D^{(\ell)}_{i_1,i_2,\cdots, i_\ell} \Phi_\ell$ is related to the Aretakis constant via
\begin{align}
A_\ell = W_{i_1,i_2,\cdots, i_\ell}(x^-) \partial_{-}D^{(\ell)}_{i_1,i_2,\cdots, i_\ell} \Phi_\ell,
\end{align}
where $W_{i_1,i_2,\cdots, i_\ell}(x^-)$ is a function of $x^-$.

We point out that $L^{(\ell+1)}_\pm$ are related to the mass annihilation operator
$D^{(\ell)}_{i_1,i_2,\cdots, i_\ell}$ in Eq.\eqref{completemassannihilation} 
up to the KGE. 
For example, $L^{(2)}_\pm$ are written as
\begin{eqnarray}
L^{(2)}_\pm &=& \pm \partial_\pm D_{-1,1} - \frac{1}{(x^+-x^-)^2}(\Box_{AdS_2}-2) \nonumber\\
          &=& \frac{1}{x^\pm}\left\{\pm\partial_\pm D_{0,1}- \frac{x^\mp}{(x^+-x^-)^2}(\Box_{AdS_2}-2)\right\} \nonumber\\
          &=& \frac{1}{(x^\pm)^2}\left\{\pm\partial_\pm D_{1,1}- \frac{(x^\mp)^2}{(x^+-x^-)^2}(\Box_{AdS_2}-2)\right\} \,,
\nonumber
\end{eqnarray}

\subsection{Aretakis constants in an extremal black hole}

We now construct the Aretakis constant in an extreme spacetime, with near horizon geometry described by $AdS_2 \times S^{n-2}$. 
We focus on a four-dimensional, extreme Reissner-Nordstr\"om geometry with unit mass.
In ingoing Eddington-Finkelstein coordinates, we have
\begin{eqnarray}
ds^2 = -\left(1 - \frac{1}{\rho}\right)^2 dv^2 + 
2 dv d\rho
+\rho^2 d\Omega^2,
\nonumber
\end{eqnarray}
with $d\Omega^2 = d\theta^2 + \sin^2\theta d\phi^2$.
Introducing $r \equiv \rho - 1$,
\begin{eqnarray}
ds^2 =
-\left[
r^2 
- r^3 \frac{(r+2)}{r+1}
\right]dv^2 + 
2  dv dr
+(r+1)^2 d\Omega^2.
\nonumber
\end{eqnarray}
The leading term in $v,r$ part is AdS$_2$ whose metric is
$ds_{AdS_2}^2 = -r^2 dv^2 + 2  dv dr$.
By using spherical harmonics on $S^2$, the massless KGE on this spacetime is written as
\begin{align}
\left[\square_{AdS_2} - \ell(\ell+1) + \partial_v 
+r K \right]\Phi = 0\,,
\label{KGeqRN0}
\end{align}
where $K$ is an operator written in the form
\begin{equation}
 K = f_1 \partial_v + f_2 + rf_3\partial_r + r^2f_4\partial_r^2 \,,
\label{eqk}
\end{equation}
with certain functions $f_1$, $f_2$, $f_3$ and $f_4$ which are regular on the horizon. 
In the near horizon limit, i.e., $v \to v/\epsilon, r \to \epsilon r$ and $\epsilon \to 0$, the equation describes
a massive scalar on AdS$_2$ with an effective mass $\ell(\ell+1)$
\begin{align}
\left[\square_{AdS_2} - \ell(\ell+1) \right]\Phi = 0.
\end{align}
However, when using the ladder operator for this spacetime, we need to 
discuss the sub-leading terms.

If we write $\Phi = e^{-r/2} \tilde{\Phi},$
then Eq.~\eqref{KGeqRN0} becomes
\begin{align}
&\left[\square_{AdS_2} - \ell(\ell+1) + r \tilde{K} \right]\tilde{\Phi} = 0,
\label{KGeqRN}
\end{align}
where $\tilde{K}$ is an operator such that $\tilde{K}\tilde{\Phi}$ is regular on 
the horizon like $K$ in Eq.\eqref{eqk}.
Next we introduce the operator $D_{i,k} \equiv {\cal L}_{\zeta_i} - k Q_i$
where $\zeta_i$ are the CCKVs on AdS$_2$ 
\begin{eqnarray}
 D_{-1,k}
&=& r^2 \partial_r + \partial_v - kr \,, \nonumber \\
 D_{0,k}
&=& r(1+vr)\partial_r + v\partial_v -k(1+vr) \,,  \nonumber \\
 D_{1,k}
&=& (v^2r^2+2vr+2)\partial_r+v^2\partial_v - kv(2+vr) \,. \nonumber 
\end{eqnarray}
Then we can see that,
\begin{align}
 \left[\square_{AdS_2} - \ell(\ell - 1) \right] {D}_{i,k}\tilde{\Phi}
=
D_{i,k-2}(\square_{AdS_2} -\ell(\ell+1))
\tilde{\Phi}
\label{RNladder1}
\end{align}
If the RHS of this equation vanishes, we can say that $D_{i,k}$ acts as a ladder operator.
However, since $\tilde{\Phi}$ satisfies Eq.\eqref{KGeqRN}, 
the RHS of this equation does not vanish.
By using Eq.\eqref{KGeqRN}, the RHS is
\begin{align}
{\rm RHS }&=
D_{i,k-2} (-r\tilde{K}\tilde{\Phi}).
\end{align}
If we choose $\zeta_{-1}$ or $\zeta_{0}$ for $\zeta_{i}$, the RHS vanishes at $r = 0$ for regular $\tilde{\Phi}$.
However, if we choose $\zeta_1$ for $\zeta_{i}$, the RHS does not vanish on the horizon because $\zeta_1$ contains $\partial_r$ with finite coefficient on the horizon. For this reason, only $D_{-1,k}$and $D_{0,k}$ can act as ladder operators.

Similar to the case of pure AdS$_2$, acting the ladder operator $\ell$ times, we can show
\begin{align}
&\square_{AdS_2} ({D}_{i_1,1}{D}_{i_2,2}\cdots {D}_{i_\ell,\ell} \tilde{\Phi})
\nonumber \\
& \quad =
D_{i_{1},-1}D_{i_{2},0} 
\cdots
D_{i_{\ell},\ell-2}(-r\tilde{K}\tilde{\Phi}).
\end{align}
If we choose $\zeta_{-1}$ or $\zeta_{0}$ for $\zeta_{i}$, RHS vanishes at $r = 0$ for regular $\Phi$.
This implies
\begin{align}
\partial_r{D}_{i_1,1}{D}_{i_2,2}\cdots {D}_{i_\ell,\ell} \tilde{\Phi} \bigg|_{r = 0} = {\rm const}
\end{align}
on the horizon because $\square_{AdS_2} \propto \partial_v\partial_r $ at $r = 0$.
If we define $A_\ell \equiv \partial_r{D}_{i_1,1}{D}_{i_2,2}\cdots {D}_{i_\ell,\ell} (e^{r/2} \Phi)$, $A_\ell$ becomes constant on the horizon.



\begin{thebibliography}{99}

\bibitem{Higuchi:1986wu}
  A.~Higuchi,
  J.\ Math.\ Phys.\  {\bf 28}, 1553 (1987)
  Erratum: [J.\ Math.\ Phys.\  {\bf 43}, 6385 (2002)].

\bibitem{Gelderen:1998}
M. van Gelderen 1998
DEOS Progress Letter 98.1 (Delft University Press, editied by R. Klees) 57-67.


\bibitem{Wald:1978vm} 
  R.~M.~Wald,
  Phys.\ Rev.\ Lett.\  {\bf 41}, 203 (1978).


\bibitem{Breitenlohner:1982jf} 
  P.~Breitenlohner and D.~Z.~Freedman,
  Annals Phys.\  {\bf 144}, 249 (1982).


\bibitem{Breitenlohner:1982bm} 
  P.~Breitenlohner and D.~Z.~Freedman,
  Phys.\ Lett.\  {\bf 115B}, 197 (1982).


\bibitem{Akhmedov:2013vka} 
  E.~T.~Akhmedov,
  Int.\ J.\ Mod.\ Phys.\ D {\bf 23}, 1430001 (2014)
  [arXiv:1309.2557 [hep-th]].


\bibitem{Cruz:1998vf} 
  J.~Cruz,
  Class.\ Quant.\ Grav.\  {\bf 16}, L23 (1999)
  [hep-th/9806145].


\bibitem{Gendenshtein:1984vs} 
  L.~E.~Gendenshtein,
  JETP Lett.\  {\bf 38}, 356 (1983)
  [Pisma Zh.\ Eksp.\ Teor.\ Fiz.\  {\bf 38}, 299 (1983)].


\bibitem{Witten:1981nf} 
  E.~Witten,
  Nucl.\ Phys.\ B {\bf 188}, 513 (1981).


\bibitem{Aretakis:2011ha} 
  S.~Aretakis,
  Commun.\ Math.\ Phys.\  {\bf 307}, 17 (2011)
  [arXiv:1110.2007 [gr-qc]].


\bibitem{Aretakis:2011hc} 
  S.~Aretakis,
  Annales Henri Poincare {\bf 12}, 1491 (2011)
  [arXiv:1110.2009 [gr-qc]].


\bibitem{Lucietti:2012xr} 
  J.~Lucietti, K.~Murata, H.~S.~Reall and N.~Tanahashi,
  JHEP {\bf 1303}, 035 (2013)
  [arXiv:1212.2557 [gr-qc]].


\bibitem{Bizon:2012we} 
  P.~Bizon and H.~Friedrich,
  Class.\ Quant.\ Grav.\  {\bf 30}, 065001 (2013)
  [arXiv:1212.0729 [gr-qc]].


\bibitem{Boyer:1976}
C. P. Boyer, E. G. Kalnins and W. Miller, Jr.,
Nagoya Math. J. 60 (1976) 35-80


\bibitem{Carter:1977pq} 
  B.~Carter,
  Phys.\ Rev.\ D {\bf 16}, 3395 (1977).


\bibitem{Eastwood:2002su} 
  M.~G.~Eastwood,
  Annals Math.\  {\bf 161}, 1645 (2005)
  [hep-th/0206233].


\bibitem{Visinescu:2010zzb} 
  M.~Visinescu,
  Europhys.\ Lett.\  {\bf 90}, 41002 (2010).


\bibitem{Batista:2014uja} 
  C.~Batista,
  Class.\ Quant.\ Grav.\  {\bf 31}, 165019 (2014)
  [arXiv:1405.4148 [gr-qc]].


\bibitem{Evnin:2015gma} 
  O.~Evnin and C.~Krishnan,
  Phys.\ Rev.\ D {\bf 91}, no. 12, 126010 (2015)
  [arXiv:1502.03749 [hep-th]].


\bibitem{Evnin:2015wyi} 
  O.~Evnin and R.~Nivesvivat,
  JHEP {\bf 1601}, 151 (2016)
  [arXiv:1512.00349 [hep-th]].


\bibitem{Evnin:2016nsz} 
  O.~Evnin and R.~Nivesvivat,
  J.\ Phys.\ A {\bf 50}, no. 1, 015202 (2017)
  [arXiv:1604.00521 [nlin.SI]].



\end{thebibliography}
\end{document}